\documentclass[aps,preprint,groupedaddress,superscriptaddress,showpacs]{revtex4-1}

\usepackage{graphicx}
\usepackage{color}
\usepackage{amsmath}
\usepackage{amsfonts}
\usepackage{amssymb}
\usepackage{dcolumn}
\usepackage{hyperref}
\begin{document}
  \title{Connectivity disruption sparks explosive epidemic spreading}

 \author{L. B\"{o}ttcher}
  \email{lucasb@ethz.ch}
  \affiliation{
 ETH Zurich, Wolfgang-Pauli-Strasse 27, CH-8093 Zurich,
Switzerland}

\author{O. Woolley-Meza}
 \affiliation{Computational Social Science, ETH Zurich, Clausiusstrasse 37, CH-8092 Zurich,
Switzerland}

\author{E. Goles}
 \affiliation{Universidad Adolfo Ib\'{a}\~{n}ez, Av. Diagonal Las Torres 2640, Pe\~{n}alol\'{e}n, Santiago, Chile}

 \author{D. Helbing}
 \affiliation{Computational Social Science, ETH Zurich, Clausiusstrasse 50, CH-8092 Zurich,
Switzerland}
   
 \author{H. J. Herrmann}
 \affiliation{
 ETH Zurich, Wolfgang-Pauli-Strasse 27, CH-8093 Zurich,
Switzerland, and Departamento de F\'isica, Universidade
Federal do Cear\'a, 60451-970 Fortaleza, Cear\'a, Brazil}

\begin{abstract}
We investigate the spread of an infection or other malfunction of cascading nature when a system component can recover only if it remains reachable from a functioning central component.
We consider the Susceptible-Infected-Susceptible (SIS) model, typical of mathematical epidemiology, on a network. Infection spreads from infected to healthy nodes, with the addition that infected nodes can only recover when they remain connected to a pre-defined \emph{central node}, through a path that contains only healthy nodes. In this system, clusters of infected nodes will absorb their non-infected interior because no path exists between the \emph{central node} and encapsulated nodes. This gives rise to the simultaneous infection of multiple nodes. Interestingly, the system converges to only one of two stationary states: either the whole population is healthy or it becomes completely infected. This simultaneous cluster infection can give rise to discontinuous jumps of different sizes in the number of failed nodes. Larger jumps emerge at lower infection rates. The network topology has an important effect on the nature of the transition: we observed hysteresis for networks with dominating local interactions. Our model shows how local spread can abruptly turn uncontrollable when it disrupts connectivity at a larger spatial scale.

\end{abstract}

\maketitle
 
\section{Introduction}
\label{sec:intro}
Spreading processes are pervasive in many fields, and have for example been analyzed in the context of ideas, rumors, behavior, and disease~
\cite{rogers2010diffusion,satorras01,kephart91,brockmann13}. One of the central insights derived from mathematical models of spreading processes, specifically disease spread, is the existence of a transition to an epidemic regime \cite{kermack27}. Using analytical and computational techniques~\cite{henkel08} from statistical physics this transition has been carefully characterized.
For example, two of the popular epidemiological models, the \emph{Susceptible-Infected-Recovered} (SIR) and the \emph{Susceptible-Infected-Susceptible} (SIS) models can be mapped to known equilibrium and non-equilibrium processes respectively \cite{satorras14,tome10,moreira96,grassberger83}.
This mapping reveals that both models exhibit a second-order phase transition, the fraction of infected or recovered nodes being the order parameter. However, under certain conditions transitions can be more abrupt, displaying first-order discontinuities \cite{herrmann15}. These transitions are of interest not only for theoretical reasons, but also because they represent unexpected and sudden changes in system behavior emerging from micro-level dynamics and are found in many naturally occurring phenomena in spreading \cite{cai15} and beyond spreading processes \cite{herrmann11,rozenfeld10,pan11}.

A recently proposed example of such discontinuous transitions is \emph{explosive percolation}. Initial work suggested that a discontinuous transition in the fraction of the largest cluster component could occur on an Erd\"os-R\'{e}nyi random network~\cite{achlioptas09}. Later work showed that this explosive percolation model, like other models using edge selection rules (\emph{product rules}), are actually continuous \cite{costa10, nagler11,grassberger11,riordan11}. However, other selection rules yield models that exhibit a first-order transition of the order parameter~\cite{araujo10,schrenk11,schrenk12,nagler12,cho13}. 

In previous work we found abrupt transitions \cite{boettcher14} when epidemic spreading induces constraints on the generation of resources needed to keep individuals healthy. However, although the transition in disease incidence can be sudden, it is nonetheless always continuous in time. Here we consider a model where healing resources always suffice but need to be distributed through pathways that are potentially obstructed through the spread of infection. Specifically, we assume that a node can only heal if there is a path between it and a \emph{central node} through healthy individuals. This represents, for example, a situation where the resources necessary for a healing process are concentrated at the \emph{central node} and only a healthy population has the capacity to keep distribution channels functioning smoothly. In general, the dynamics apply to any system where proper function requires connectivity to a providing node or power source, such as an electric power grid. This induces the possibility that healthy nodes become encapsulated by infected ones and thus are absorbed into an infected cluster all at once. Interestingly, we find and prove rigorously that these dynamics only allow two possible fixed points: a fully infected and a fully healthy state. Furthermore, we observe hysteresis when the network topology is dominated by local connections. The encapsulation dynamics of the cluster infection also resemble biological processes such as is the tubercle formation in lungs \cite{russel-tubercle}. Finally, the model can represent information dynamics in a social system, for example where individuals who are disconnected from a source of true information can be trapped in a bubble of misinformation.
\section{Model and Methods}
\label{sec:model}
The underlying spreading model we choose is an SIS model since it is the simplest epidemiological model which captures long time scales. Under the SIS rules individuals are either in an infected or a susceptible state. Susceptible nodes become infected at rate $p$ through contact with their infected nearest neighbors. Infected nodes heal at rate $q$. This process can be illustrated by the following reaction schema:
\begin{equation*}
S\overset{p}{\rightarrow} I \overset{q}{\rightarrow} S.
\end{equation*}
We define $\tau= \langle k \rangle p/q$ for a network with average degree $\langle k \rangle$ as a measure of the spreading effect and a critical value $\tau_c$ above which the epidemic will break out. Note that in a mean-field SIS model, $\tau$ is the basic reproduction number.
\begin{figure}
\centering
\includegraphics[width=0.75\textwidth]{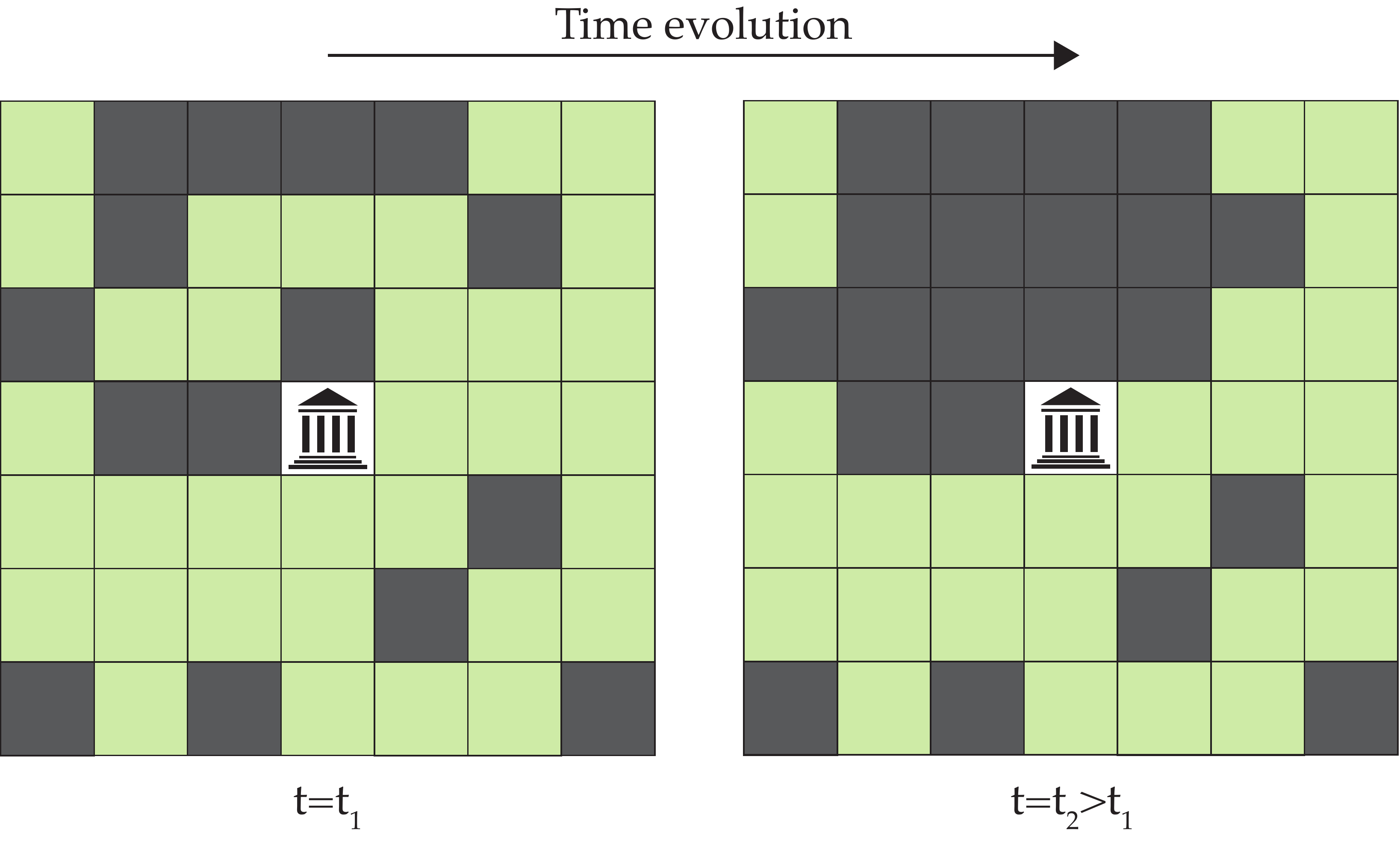}
\caption{\textbf{Schematic illustration of the cluster SIS model.} In the cluster SIS model all healing process are controlled by the central node (government symbol). The healing of an infected node occurs if, and only if, a path of healthy sites (light green) connects the central node with the infected one (dark grey). At every time step we check if infectious nodes surround healthy ones. If yes, this leads to the cluster infection process of the inner healthy nodes, shown at time $t_2$. In this example the node above the central node infects its right neighbor. However, five additional nodes are now encapsulated and will become infected too.}
\label{fig:clustersis_sketch}
\end{figure}

Our \emph{cluster SIS} model assumes a modified update mechanism (see Fig.~\ref{fig:clustersis_sketch}). Specifically, there is a single central node in the system that controls healing. An infected node can heal if, and only if, it is connected to the \emph{central node} via a path involving only healthy nodes. When infected lattice sites encapsulate healthy ones, all of the inner healthy subpopulation becomes immediately absorbed in the surrounding infected cluster, since they are cut-off from the central node. At some point the central node may also become surrounded by infected nodes, leading to a sudden jump in infection that soon leads to a fully infected absorbing state.
This clustering mechanism resembles the update schema in the recently introduced \emph{no-enclave percolation} (NEP) model \cite{sheinman15}. 

We simulate the cluster SIS model using a kinetic Monte Carlo method (Gillespie's algorithm) \cite{gillespie76,gillespie77}. Recovery with rate $q$ and infection with rate $p$ are the two processes defining the dynamics of the system. After an infection event certain regions might have lost their connectivity to the central component. We apply the burning algorithm~\cite{herrmann84,stauffer-aharony85} to determine these encapsulated nodes and then infect them all at once. Healing processes require connectivity to the central component and thus only occur at the perimeter of the encapsulated regions. A successful recovery event re-establishes connectivity between the previously encapsulated nearest-neighbors of the healed node and the central component.

In our analysis we focus on a square lattice, but we will also briefly address the dynamics on a school friendship network and an Apollonian graph (see \emph{Additional Information} and \emph{Appendix}). The square lattice only accounts for local, nearest-neighbor interactions. However, the school network possesses additional topological features, such as long-range connections and community structure, which affect the spreading process. 
The Apollonian graph shows scale-free, small world and matching graph properties \cite{andrade05}. For the two latter networks the node with the largest degree is chosen as the \emph{central node}.
\section{Results}
\label{sec:results}
\subsection{Discontinuous clustering}
\begin{figure}
\centering
\includegraphics[width=1\textwidth]{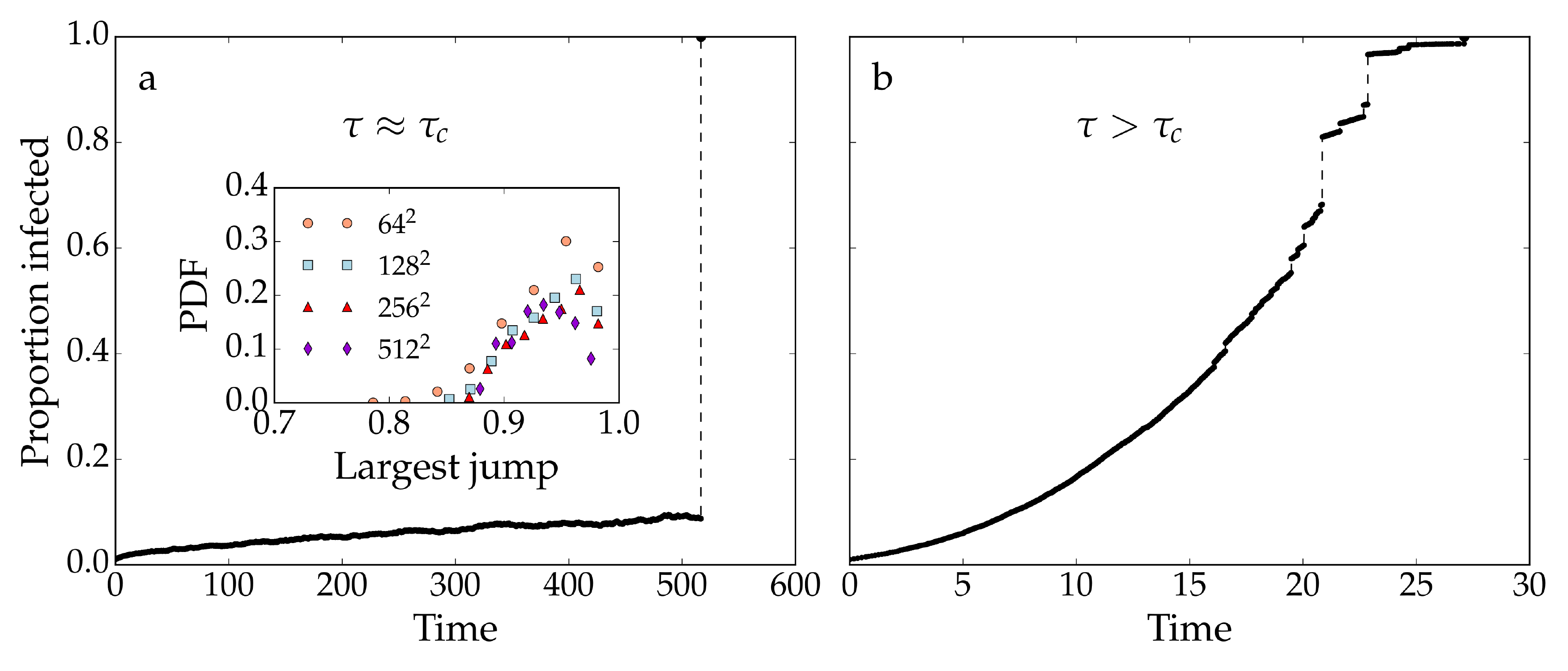}
\caption{\textbf{Discontinuous behavior in time.} Proportion of infected nodes as a function of time for the cluster SIS model with $q=0.4$, $k=4.0$ on a $128\times128$ square lattice. \textbf{(a)} Close above the threshold $\tau_c$ with $p=0.165$ one encounters large jumps (surrounded central node). The inset shows the distribution of largest jumps (binned data) for $L\times L$ square lattices with $L=64,~128,~256$ ($2\cdot 10^3$ samples) and $L=512$ (500 samples). \textbf{(b)} For $\tau>\tau_c$ ($p=0.3$) one also observes smaller jumps which correspond to a large number of smaller encapsulated regions.}
\label{fig:cSIS_jumps}
\end{figure}

One important characteristic of the cluster SIS model is the possibility of updating multiple lattice sites at once. As a consequence, one observes jumps of different sizes in the time evolution of the fraction of infected individuals. For different values of the control parameter $\tau$ the time evolution on a square lattice is illustrated in Fig.~\ref{fig:cSIS_jumps}. 
The critical value $\tau_c$ is determined by the underlying contact process dynamics of the simple SIS model. The cluster update processes only become important in the epidemic regime $\tau>\tau_c$ (for the square lattice $\tau_c=1.6488(1)$ \cite{moreira96}). For $\tau$ close to the threshold value $\tau_c$, the epidemic starts to slowly grow (Fig.~\ref{fig:cSIS_jumps}a). Close to $\tau_c$ we observe a qualitatively different behavior compared to larger values of $\tau$ (Fig.~\ref{fig:cSIS_jumps}b). Namely, in the vicinity of $\tau_c$ the time evolution of the proportion of infected nodes exhibits a very large jump (the distribution of the largest jump for different system sizes is shown in the inset of Fig.~\ref{fig:cSIS_jumps}a). This jump corresponds to a slowly spreading epidemic encapsulating a large fraction of all nodes by surrounding the center node after a certain time. For $\tau$ larger than $\tau_c$ the time evolution involves a higher number of smaller jumps due to the existence of more encapsulated regions (Fig.~\ref{fig:cSIS_jumps}b). A more detailed analysis of the jump size distribution is presented in Figs. \ref{fig:jump_distr_sl} and \ref{fig:jump_distr_sl_fit} (see \emph{Appendix}). As expected, the characteristic time after which the stationary state has been reached is smaller for larger $\tau$ and as long as the infection rate of the disease leads to an epidemic ($\tau\ge\tau_c$), abrupt jumps in infection levels are much more likely when the transmissibility is lower.
\subsection{Transition time}
\begin{figure}
\centering
\includegraphics[width=0.65\textwidth]{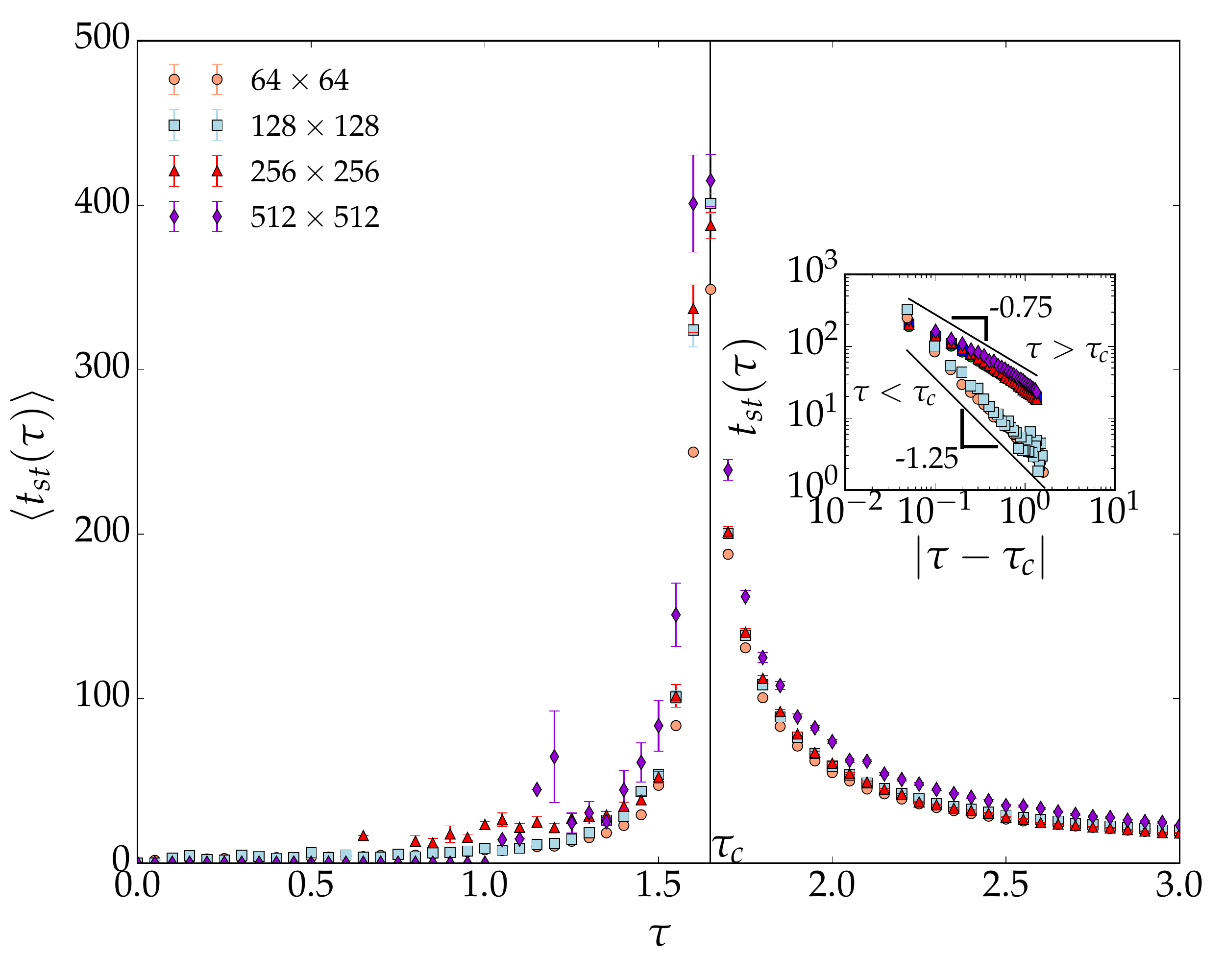}
\caption{\textbf{Transition time to the fully infected state for square lattices---strongly increasing close to $\tau_c$.} Transition time $\langle t_{st}(\tau) \rangle$ to fully infected state as function of $\tau_c$ for different $L\times L$ square lattices $L=64,~128,~256,~512$ ($2\cdot 10^4$, $4\cdot 10^3$, $2\cdot 10^3$ and 500 samples). The inset shows both regimes $\tau<\tau_c$ (only $L=64,~128$) and $\tau>\tau_c$ on a double-logarithmic scale. The straight lines in the inset are guides to the eye with slopes $-1.25$ for $\tau<\tau_c$ and $-0.75$ for $\tau>\tau_c$. Error bars indicate the standard error of the mean.}
\label{fig:cSIS_tst}
\end{figure}
The time needed until the fully infected regime is reached will be referred to as $t_{st}(\tau)$ for a single realization and $\langle t_{st}(\tau) \rangle$ for the averaged value. An illustration of the time dependence is given in Fig.~\ref{fig:cSIS_jumps}. Close to $\tau_c$ the system needs much more time to reach the stationary state compared to $\tau\gg\tau_c$. In other words, the system exhibits \emph{critical slowing down}. A detailed analysis of this effect on the square lattice is presented in Fig.~\ref{fig:cSIS_tst}. Even for $\tau<\tau_c$ we found a few samples which converge to a fully infected population. For example, this might occur when the initial infection occurs close to the central node. The inset of Fig.~\ref{fig:cSIS_tst} shows both regimes $\tau<\tau_c$ and $\tau>\tau_c$ in a double-logarithmic scale.

\subsection{Emergence of a first order phase transition}
\begin{figure}
\centering
\includegraphics[width=0.65\textwidth]{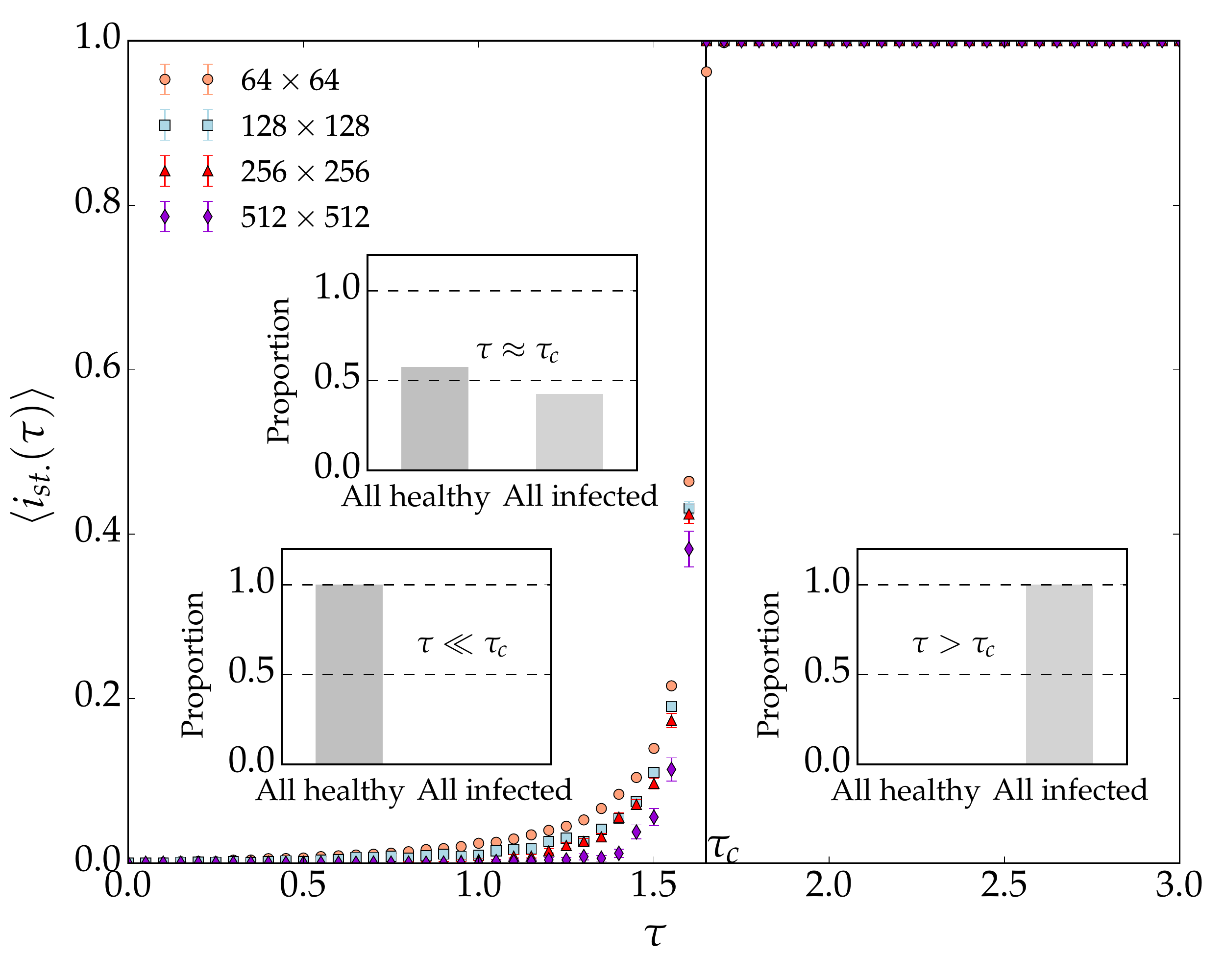}
\caption{\textbf{Stationary state with a jump close to $\tau_c$.} The order parameter $\langle i_{st}(\tau)\rangle$ as a function of $\tau$ for different $L\times L$ square lattices $L=64,~128,~256,~512$ ($2\cdot 10^4$, $4\cdot 10^3$, $2\cdot 10^3$ and 500 samples). Above the threshold $\tau_c$ only fully infected states are reached. Error bars indicate the standard error of the mean. Insets show the proportions of the two possible states for different values of $\tau$. Close to $\tau_c$ the distribution is bimodal, indicating a discontinuous transition.}
\label{fig:cSIS_ist}
\end{figure}

After a certain time our cluster SIS model reaches its stationary state. The transition of the stationary state as a function of $\tau$ can be characterized by the fraction of infected nodes. For a single realization we will refer to the corresponding fraction as $i_{st}(\tau)$, whereas the averaged quantity $\langle i_{st}(\tau)\rangle$ is interpreted as the order parameter.
Fig.~\ref{fig:cSIS_ist} shows how this order parameter changes on a square lattice for different values of $\tau$. An important characteristic of $i_{st}(\tau)$ is the occurrence of only two possible values, namely a healthy and a completely infected population, i.e. $i_{st}(\tau)\in\{0,1\}$. In order to prove that these are the only two possible attractors of the system, we can use a proof by contradiction. Consider a connected graph with $N$ nodes $x_i\in\{0,1\}$ where $i\in\{1,\dots,N\}$ and assume the existence of another fixed point consisting of a mixture of healthy and infected nodes. Thus, there would exist a pair of nodes $x_i=0$ and $x_j=1$ for $i\neq j$. If the central node is infected, the stationary state would correspond to a fully infected population in contradiction to the assumption. Thus, the central node has to (1) be healthy and (2) have at least one healthy neighbor. Furthermore, there exists at least one pair that contains an infected node $x_j$ and a neighboring healthy node. From this fact, it follows directly that the healthy node next to the infected one $x_j$ can become infected too. There must also exist a path of healthy nodes to the infected node $x_j$, and thus it can become healthy. Consequently, the configuration cannot be an attracting fixed point and we have a contradiction. We therefore conclude that the dynamics only allows a fully infected or healthy fixed point. 
\begin{figure}
\begin{minipage}[b]{0.49\textwidth}
\centering
\includegraphics[width=\textwidth]{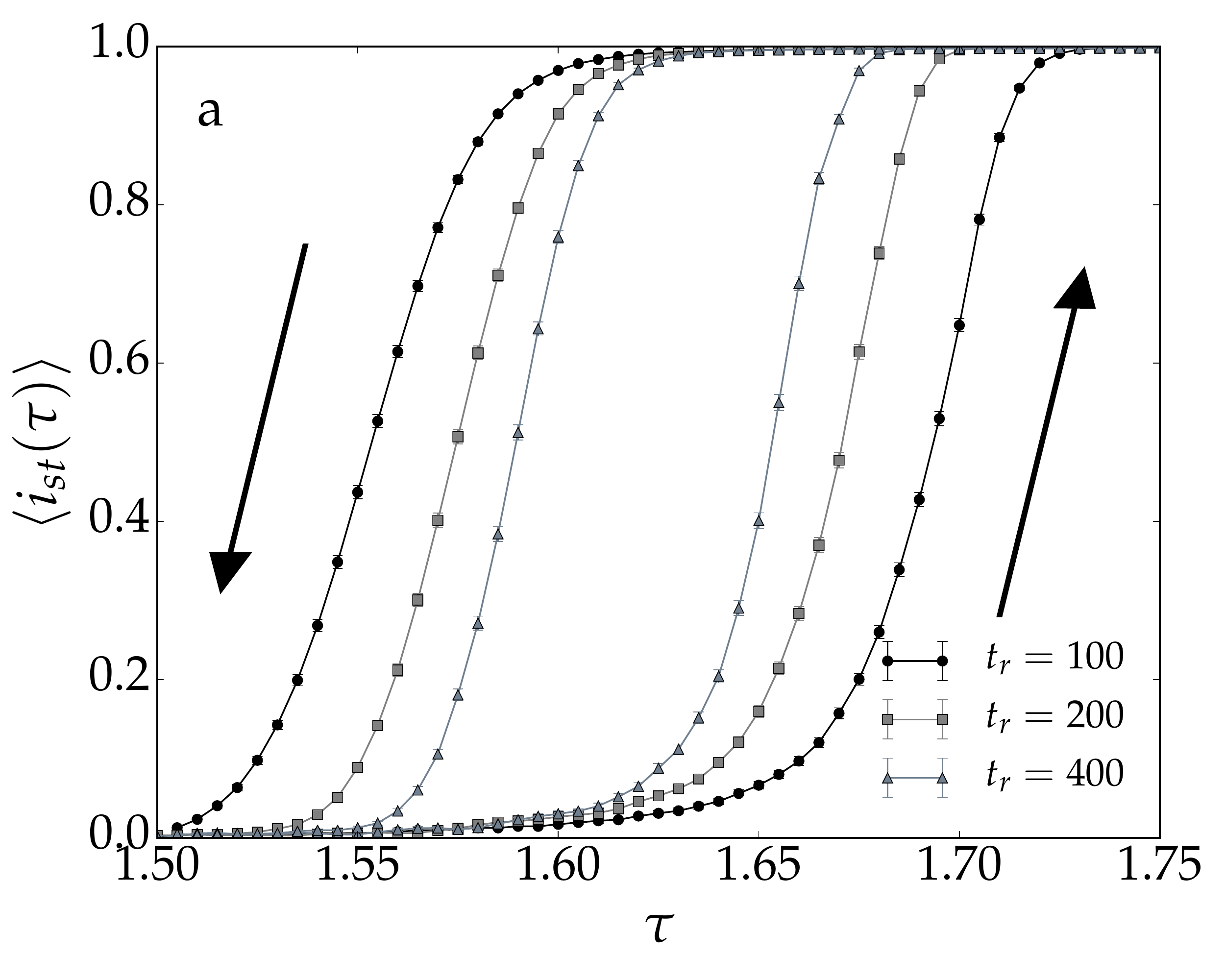}
\end{minipage}
\hspace*{\fill} % separation between the subfigures
\begin{minipage}[b]{0.49\textwidth}
\centering
\includegraphics[width=\textwidth]{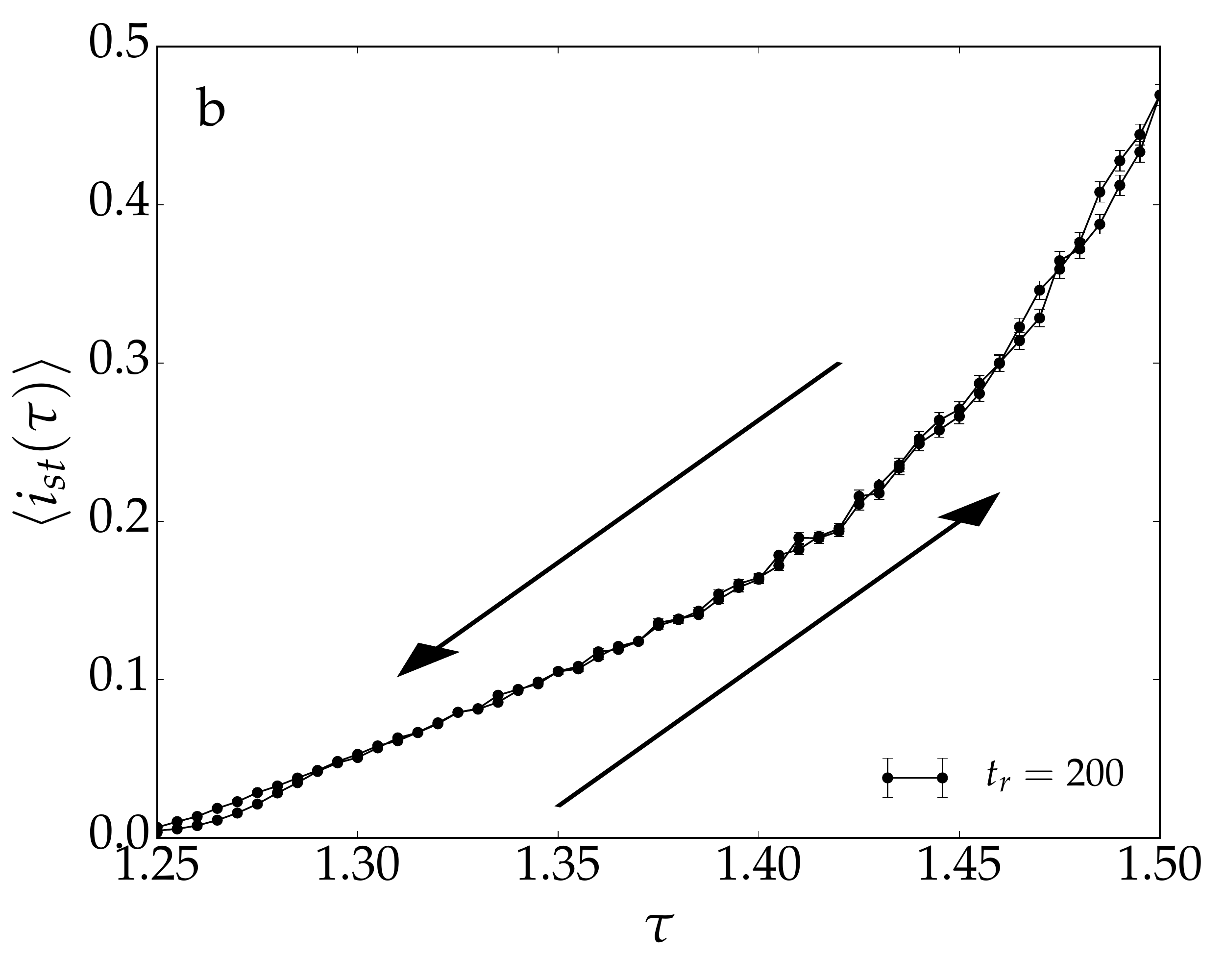}
\end{minipage}
\hspace*{\fill} % separation between the subfigures
  \caption{\textbf{Hysteresis study on the square lattice without and with long-range connections.} The hysteresis effect of the order parameter $\langle i_{st}(\tau)\rangle$ as a function of $\tau$ on a \textbf{(a)} $128 \times 128$ square lattice with different relaxation times $t_r=100,~200,~400$ ($2\cdot 10^3$ samples) and \textbf{(b)} on a $128\times128$ square lattice with long-range connections and $t_r=200$ ($r=0.5$, $\langle k \rangle=4.99$ and $10^3$ samples). The SCM has been applied to both networks assuming a minimum fraction of active sites $\epsilon=0.0005$. In the network with additional long-range connections the probability $r$ defines the average number of additional undirected links per node. Arrows indicate the direction of the hysteresis loop and error bars indicate the standard error of the mean.}
 \label{fig:hyst}
\end{figure}

We expect an abrupt transition from the fully healthy to the fully infected phase. This rapid change of the order parameter implies a first-order transition. Further evidence of the discontinuity is provided by the bimodal distribution of $\langle i_{st}(\tau)\rangle$ close to $\tau_c$ (see insets of Fig.~\ref{fig:cSIS_ist}) and the collapse of the order parameter $\langle i_{st}(\tau)\rangle$ onto a single curve for $\tau < \tau_c$, suggesting that it is approaching zero in the limit of an infinite system (see Fig.~\ref{fig:cSIS_scaling} in the \emph{Appendix}). Finite size effects smoothen out the discontinuity, and although the transition of $i_{st}(\tau)$ is discontinuous for every single realization, averaged over multiple realizations, the transition of the order parameter $\langle i_{st}(\tau)\rangle$ may appear smeared out due to finite system sizes as in Fig. \ref{fig:cSIS_ist}.

To further investigate the transition we apply the \emph{spontaneous creation method} (SCM) to the system \cite{bidaux89}. This method is used to study the hysteresis effect accompanying first-order transitions \cite{dickman91,monetti01}. For non-equilibrium processes one is typically confronted with an absorbing and a fluctuating state \cite{henkel08}. An absorbing state can be reached by the dynamics, but not left anymore. The SCM assumes a small fraction $\epsilon$ of active nodes which also survives in the absorbing completely healthy state. This turns the absorbing state into a fluctuating one with average density $\epsilon$. A second-order phase transition will not survive this modification, whereas a first-order transition will display hysteresis between a low- and high-density phase. We apply the SCM to a square lattice with a minimum fraction of active sites $\epsilon=0.0005$. In addition, we do not allow the central node to become infected in order to avoid getting trapped in the completely infected state. It is clear from Fig.~\ref{fig:hyst}a that the system shows hysteresis for different relaxation times $t_r$, which have been chosen according to the characteristic time-scale in Fig.~\ref{fig:cSIS_tst}.
\subsection{The effects of topology on the transition}

How robust is this transition behavior to changes in network topologies? To address this question we briefly investigate the dynamics on the school friendship and the Apollonian networks (see Fig.~\ref{fig:cSIS_hyst_school_appol} in the \emph{Appendix}).
In fact, performing the same SCM analysis as above, with $\epsilon=0.005$ and relaxation times according to the characteristic time scale we find no hysteresis for the school friendship network, in contrast to the square lattice. 

One possibility would be that the hysteresis effect accompanying first order transitions only occurs when local connectivity dominates the topology. We test this claim by adding long-range connections to the square lattice. 
With probability $r$ a node gets a new connection to another, randomly selected, node.
We then apply the SCM to the square lattice with long-range connections and $r=0.5$. As illustrated in Fig.~\ref{fig:hyst}b, no hysteresis is observable, giving support to the idea that the first order transition is destroyed by long-range interactions. This makes sense intuitively since an increasing fraction of
long-range connections approaches the mean-field limit. In the mean-field description the central node would be connected to any node in the network. Thus, the appearance of an abrupt transition strongly depends on the underlying topology.
We note, however, that we have no proof that the first-order transition necessarily disappears with long-range connections. The situation may be more subtle. In fact, in the \emph{Appendix} we show that for an Apollonian network (which has long-range connections) the SCM still yields hysteresis behavior (Fig. \ref{fig:cSIS_hyst_school_appol}b). Thus, a further investigation of the effect of different long-range connection or central node densities is one possible direction for future work.

\section{Discussion}
\label{sec:discussion}
We explore epidemic dynamics under the constraint that a node can only heal if healing resources localized at a \emph{central node} are reachable through a path of healthy nodes. In such a model, instead of infecting only one node per time step, a closed infected cluster will instantaneously absorb its non-infected interior since it is no longer connected to the \emph{central node}. With this cluster update mechanism the system exhibits only two attractors: either the whole system ends up healthy or completely infected. Since the cluster dynamics becomes relevant only once when the epidemic starts growing, the threshold $\tau_c$ is the same as for the SIS contact process. In a single realization the level of infection in the system discontinuously jumps from zero to one. Averaging over multiple realizations leads to a smeared-out transition of the order parameter depending on the system size. However, we show that on a square lattice this transition is still first-order. 
Furthermore, we also find that the topology strongly influences the cluster dynamics. Specifically, with the addition of long-range connections to the lattice, hysteresis cannot be observed anymore. Future work might further investigate this effect in terms of a parametric study of varying central node and long-range connection densities.
Intuitively, in the mean-field description the \emph{central node} is connected to every node in the network. Adding long-range connections to the topology approaches this mean-field limit and thus simple SIS dynamics without a discontinuous jump.

This model very generally captures the unexpected dynamics that can emerge when failure spreads locally and can only be repaired as long as the damage does not block global connectivity.
The intuitive approach of lowering the transmissibility of failure to make a system more robust could have the unintended consequence of taking the system to a regime where global failure can happen more suddenly. 
The insights we derive here are applicable to many different socio-technical systems, such as road networks, power grids or the dynamics of socially transmitted information such as opinions or sentiment.

\section*{Acknowledgments}
We want to thank Jos\'e Soares Andrade J\'unior for the Apollonian network data. We acknowledge financial support from the ETH Risk Center and ERC Advanced grant number FP7-319968 FlowCCS of the European Research Council. E.G. would like to thank H.J.H. for his invitation to Zurich and acknowledges financial support by the grants Fondecyt-Chile-1140090 and Milenio NS130017. This work was partially funded by the European Community's H2020 Program under the funding scheme „FETPROACT-1-2014: Global Systems Science (GSS)”, grant agreement \#641191 „CIMPLEX: Bringing CItizens, Models and Data together in Participatory, Interactive SociaL EXploratories” (http://www.cimplex-project.eu).
\section*{Author contribution statement}
H.J.H., O.W-M., and D.H. conceived and supervised the study, L.B. carried out
computational simulations and E.G. formulated the proof. L.B., O.W-M. and E.G. wrote the manuscript. All authors reviewed the manuscript.
\section*{Additional information}
\textbf{Competing financial interests} The author(s) declare no competing financial interests.
\textbf{Data sources} This research uses the public-use dataset from Add Health, a program project
designed by J. Richard Udry, Peter S. Bearman, and Kathleen Mullan Harris, and funded
by a grant from the National Institute of Child Health and Human Development (P01-HD31921). 
For data files from Add Health contact Add Health, Carolina Population Center,123 W. Franklin Street, Chapel Hill, NC 27516-2524, http://www.cpc.unc.edu/addhealth.
\clearpage
\appendix
\section{Topologies}
\begin{figure}
\begin{minipage}[b]{0.49\textwidth}
\centering
\includegraphics[height=5.5cm]{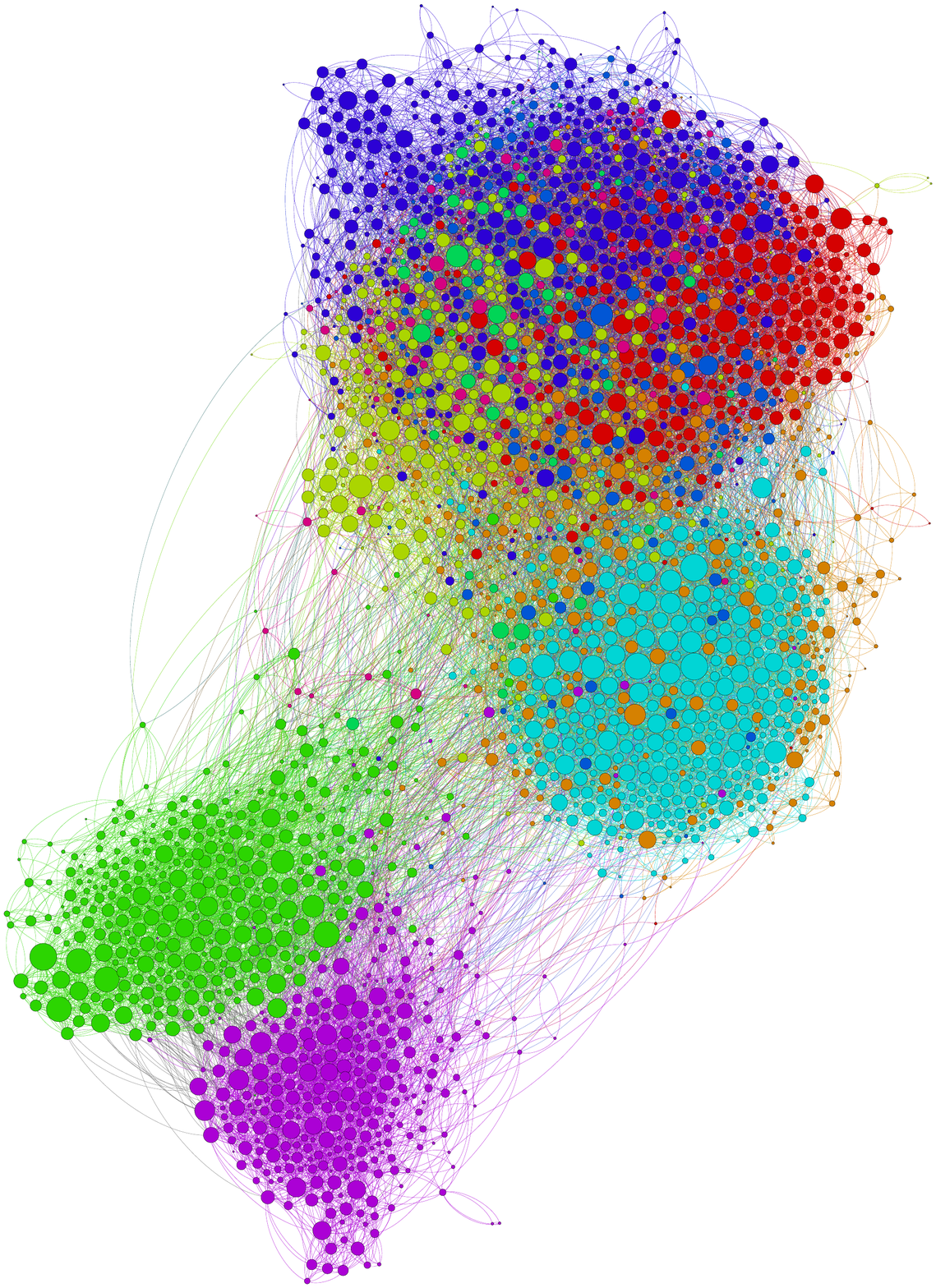}
\caption{School network (Add Health) with 2539 nodes and average degree $\langle k \rangle \approx 8.24$ (created with Gephi \cite{gephi09}).} \label{fig:school_gephi}
\end{minipage}
\hspace*{\fill} % separation between the subfigures
\begin{minipage}[b]{0.49\textwidth}
\centering
\includegraphics[height=5.5cm]{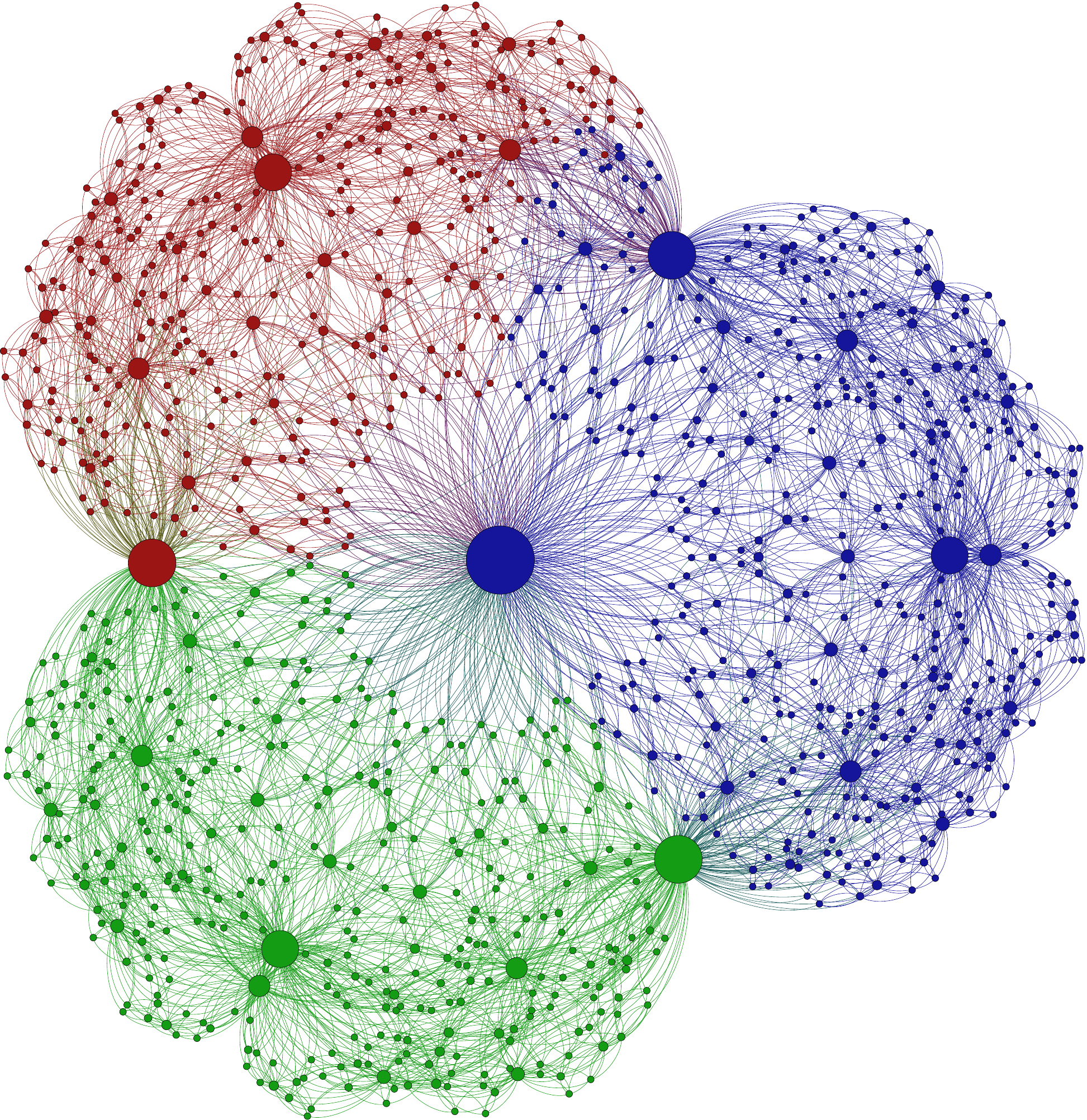}
\caption{Apollonian network with 1096 nodes and average degree $\langle k \rangle \approx 5.99$ (created with Gephi \cite{gephi09}).} \label{fig:apollonian_gephi}
\end{minipage}
\hspace*{\fill} % separation between the subfigures
\end{figure}

In addition to the square lattice we consider a school friendship and an Apollonian network. The cluster spreading model is characterized by a central node which is responsible for the healing process. In particular, the centralized character applies for Apollonian networks where onion-like structures are all connected to a central node as illustrated in Fig.~\ref{fig:apollonian_gephi}. The sixth-generation Apollonian network with 1096 nodes and average degree $\langle k \rangle\approx 5.99$ is used in this study.
The school network involves long-range connections, clustering and community structure as shown in Fig.~\ref{fig:school_gephi}. In this study only the largest school network of the Add Health program with 2539 nodes and an average degree of $\langle k \rangle\approx 8.24$ will be considered.
\section{Scaling and hysteresis effects}
\begin{figure}
\centering
\includegraphics[width=0.65\textwidth]{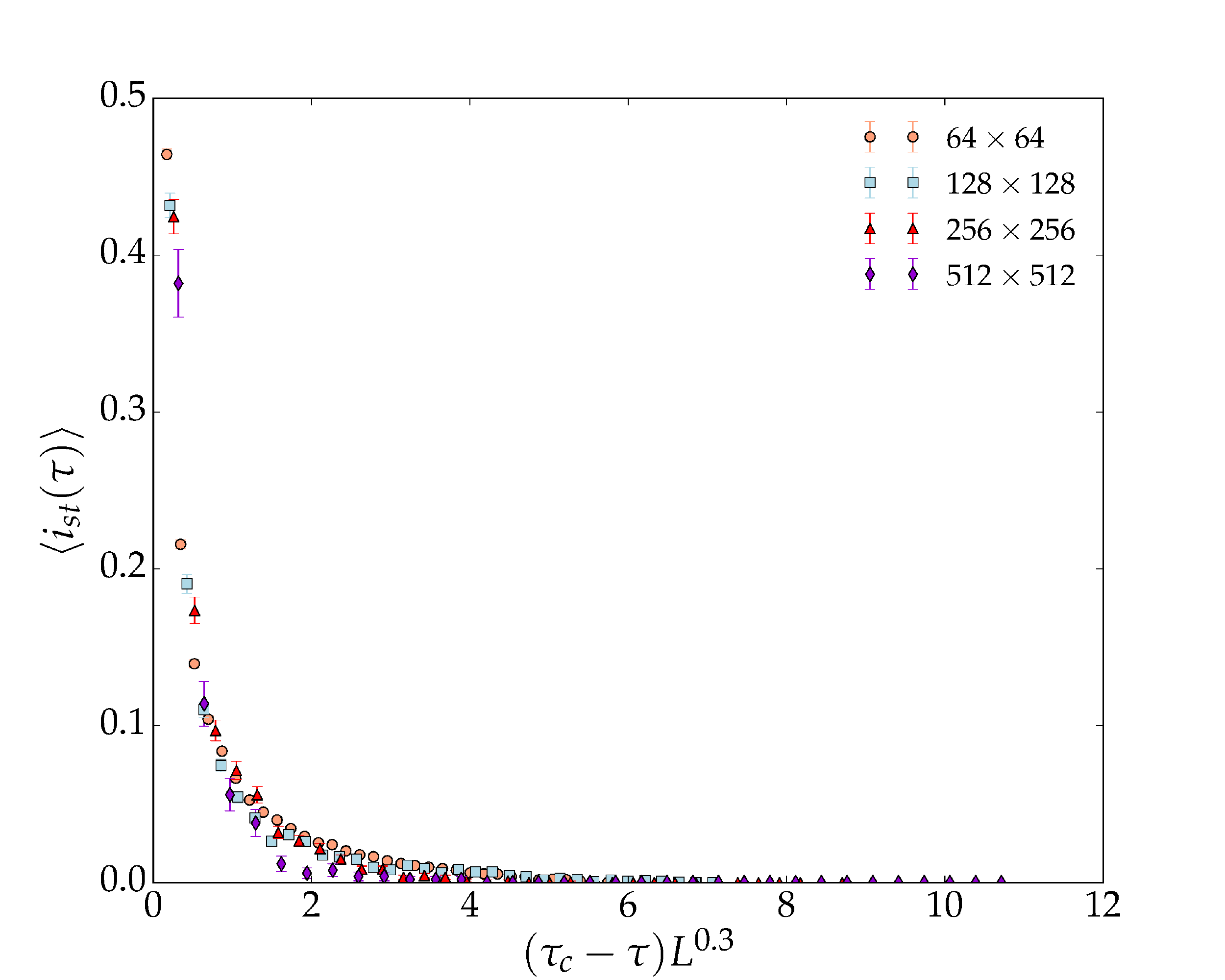}
\caption{\textbf{Data collapse of the order parameter for different system sizes.} The order parameter $\langle i_{st}(\tau) \rangle$ as a function of $(\tau_c-\tau) L^{0.3}$ for $\tau < \tau_c$ and different $L\times L$ square lattices $L=64,~128,~256,~512$ ($2\cdot 10^4$, $4\cdot 10^3$, $2\cdot 10^3$ and 500 samples). The data collapse onto a single curve suggesting that the order parameter approaches zero in the limit of an infinite system. Error bars indicate the standard error of the mean.}
\label{fig:cSIS_scaling}
\end{figure}

\begin{figure}
\begin{minipage}[b]{0.49\textwidth}
\centering
\includegraphics[width=\textwidth]{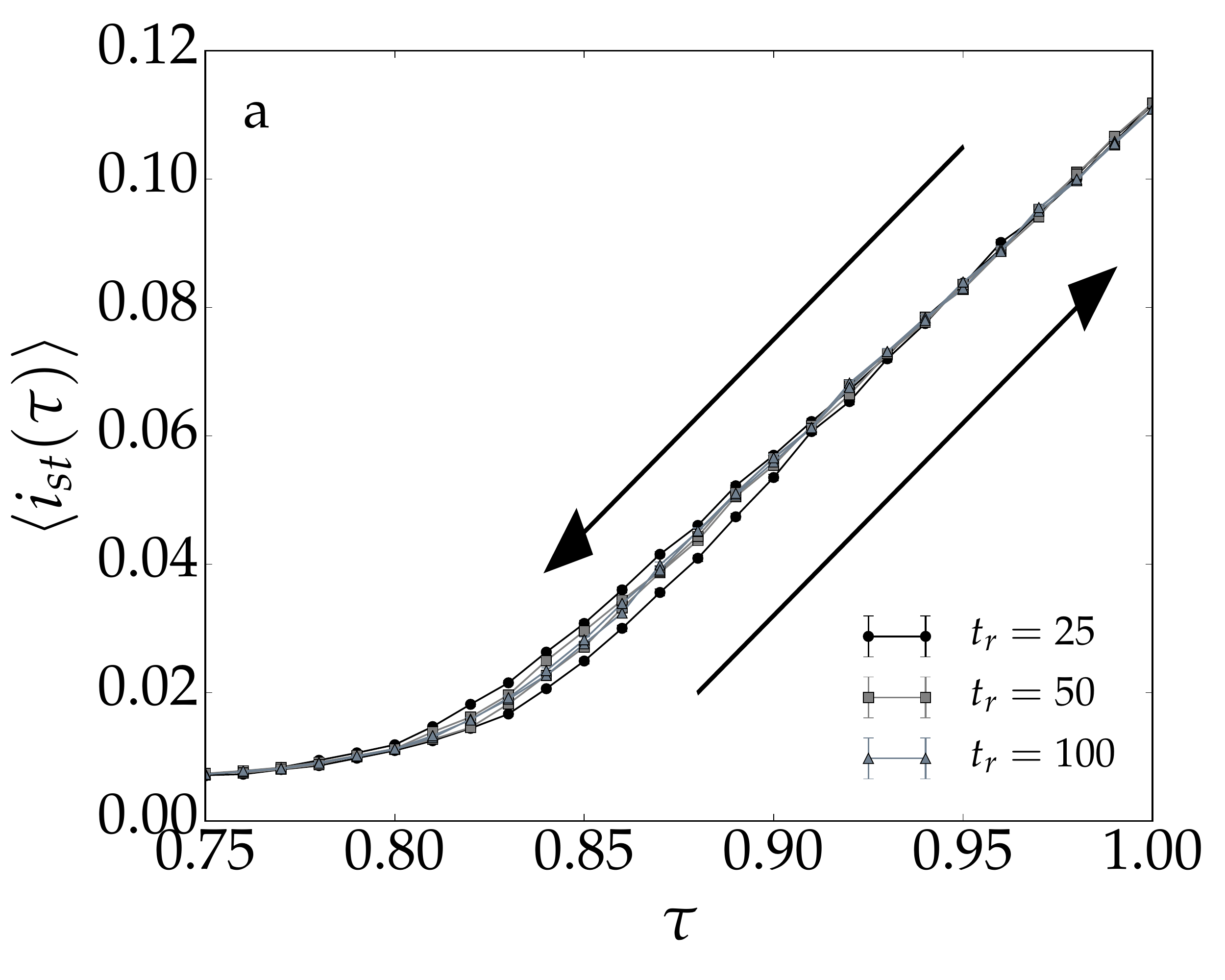}
\end{minipage}
\hspace*{\fill} % separation between the subfigures
\begin{minipage}[b]{0.49\textwidth}
\centering
\includegraphics[width=\textwidth]{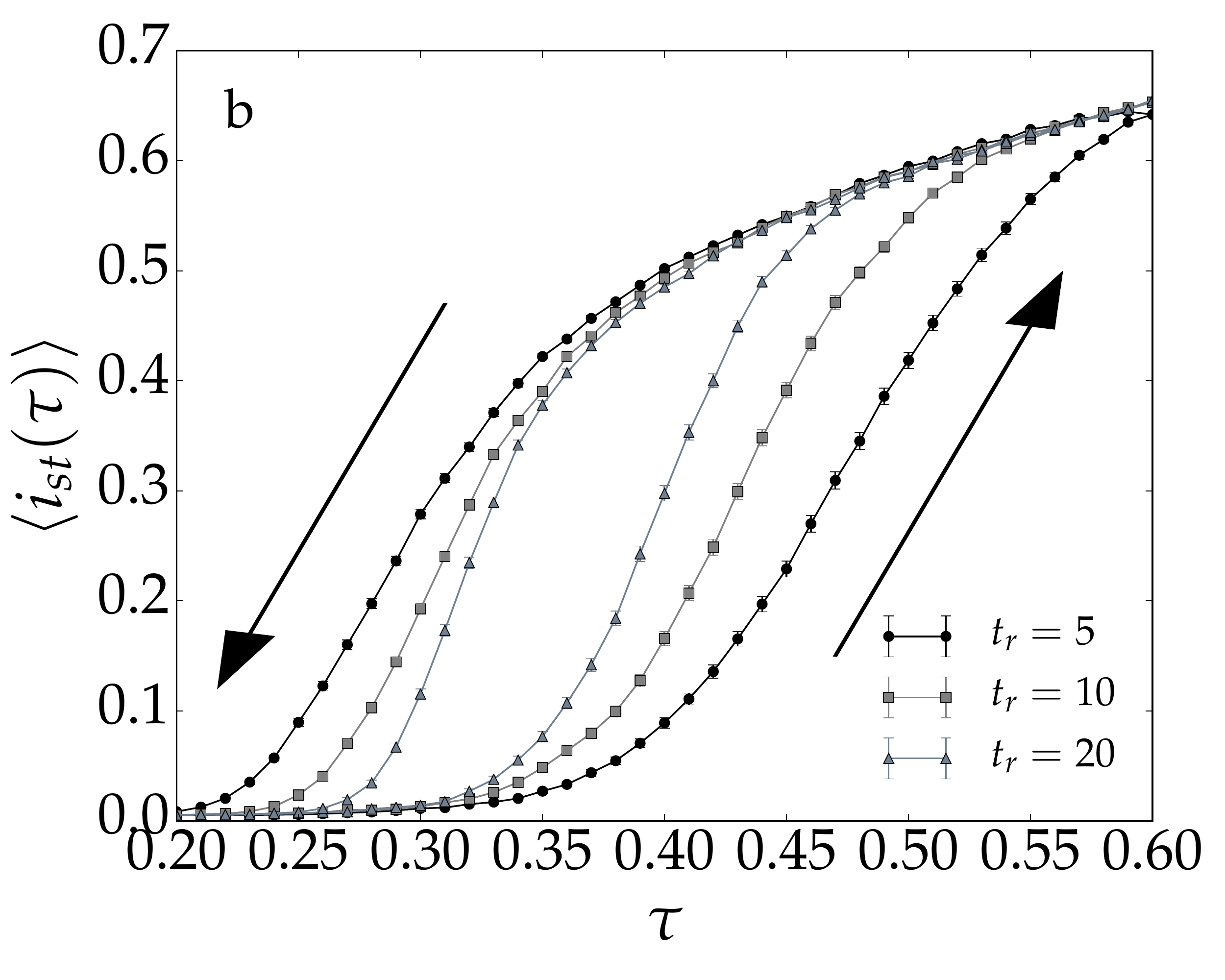}
\end{minipage}
\hspace*{\fill} % separation between the subfigures
  \caption{\textbf{Hysteresis curves for the school friendship and the Apollonian network.} The hysteresis effect of the order parameter $\langle i_{st}(\tau)\rangle$ as a function of $\tau$ for the \textbf{(a)} school network and \textbf{(b)} Apollonian network (both $10^3$ samples). Different relaxation times $t_r$ have been studied with the SCM and a minimum fraction of active sites $\epsilon=0.005$. The arrows indicate the direction of the hysteresis loop and the error bars indicate the standard error of the mean. In contrast to the school network the Apollonian network shows hysteresis.}
   \label{fig:cSIS_hyst_school_appol}
\end{figure}

In order to support the claimed first-order transition on the square lattice, Fig. \ref{fig:cSIS_scaling} shows a data collapse of the order parameter $\langle i_{st}(\tau) \rangle$ as a function of $(\tau_c-\tau) L^{0.3}$ for $\tau < \tau_c$. This suggests that the order parameter approaches zero for $\tau < \tau_c$ in the limit of an infinite system. As described in the main text the hysteresis effect has been also studied for the school friendship and the Apollonian networks by applying the SCM and assuming a minimum of fraction of active sites $\epsilon=0.005$. In addition, the central node is not allowed to become infected in order not to get trapped in the fully infected state. The results are shown in Figs. \ref{fig:cSIS_hyst_school_appol}a and \ref{fig:cSIS_hyst_school_appol}b. Similar to the square lattice with long-range connections the school network shows no hysteresis. However, regarding the Apollonian network the situation is more subtle. Although the Apollonian network also shows small-world properties, the SCM still exhibits hysteresis.
\section{Jump size distribution}
\label{sec:jump_size_distr}
\begin{figure}
\centering
\includegraphics[width=1\textwidth]{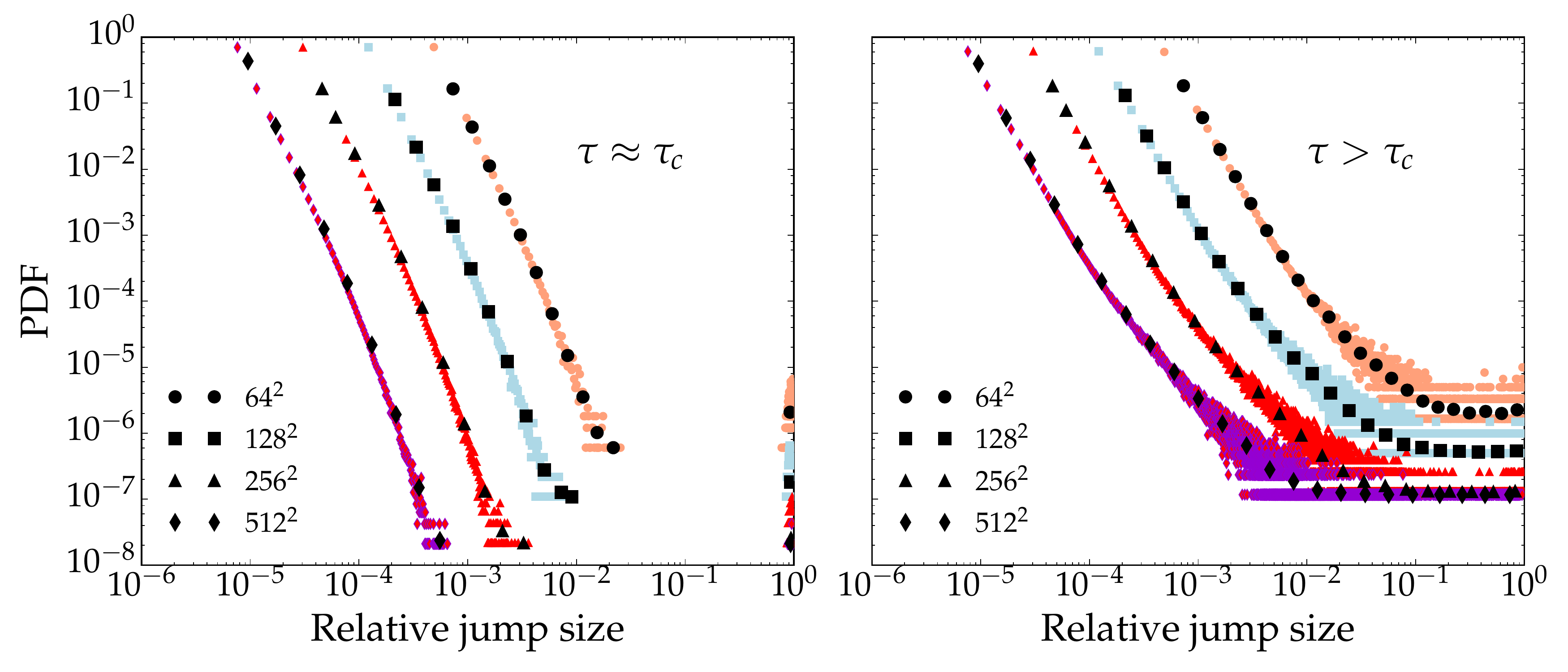}
\caption{\textbf{Distribution of relative jump sizes for $\tau=1.65,~3.00$.} The relative jump size distribution for different $L\times L$ square lattices with $L=64,~128,~256$ ($2\cdot 10^3$ samples) and $L=512$ (500 samples). All jumps are normalized by the corresponding system sizes. The distribution starts with the smallest possible jump of size two. For $\tau=1.65\approx\tau_c$ the distribution has two separated regions. The separation vanishes for larger $\tau>\tau_c$.}
\label{fig:jump_distr_sl}
\end{figure}
\begin{figure}
\centering
\includegraphics[width=0.65\textwidth]{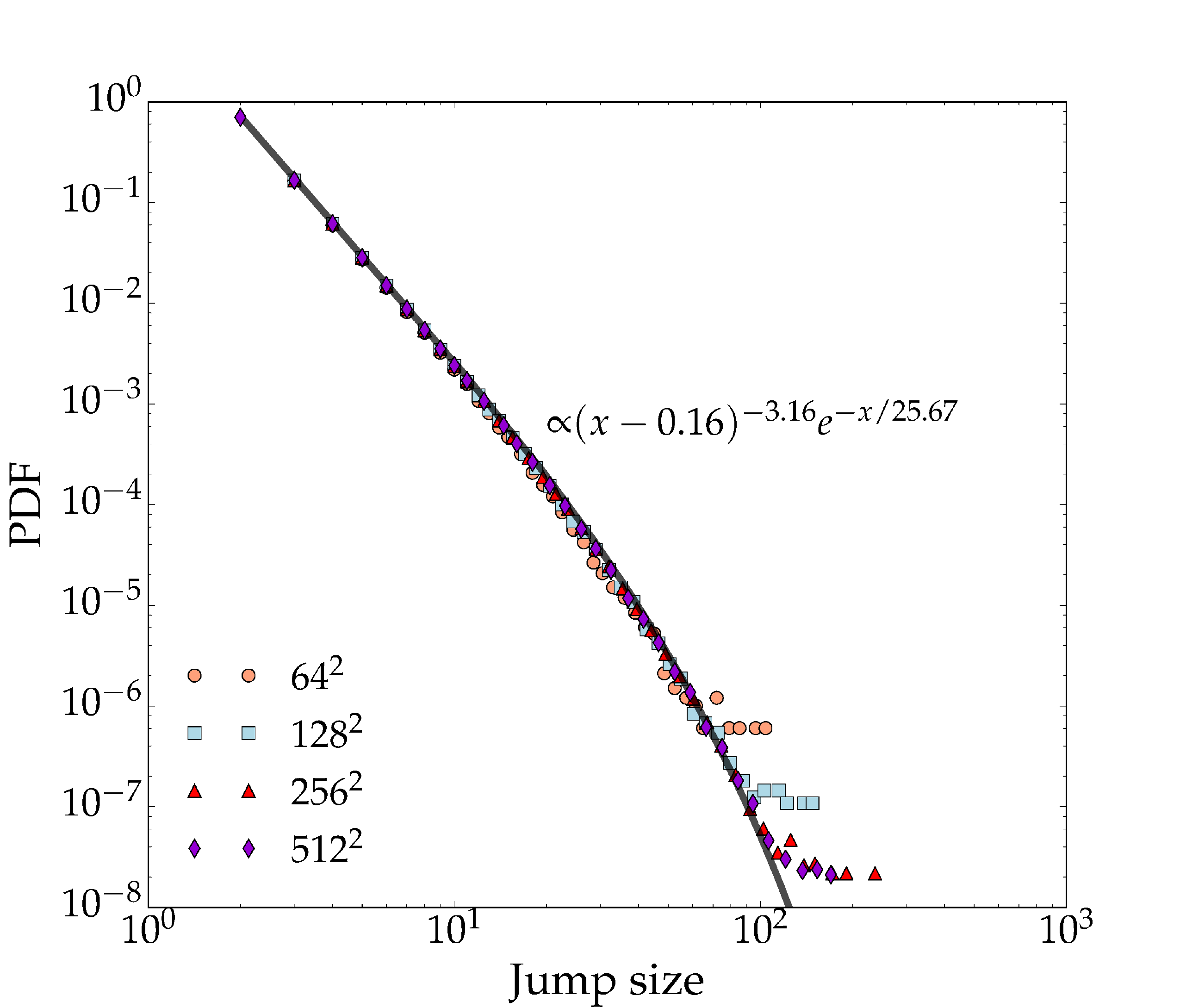}
\caption{\textbf{Distribution of jump sizes for $\tau=1.65\approx \tau_c$.} The jump size distribution for different $L\times L$ square lattices with $L=64,~128,~256$ ($2\cdot 10^3$ samples) and $L=512$ (500 samples). The curves correspond to the ones from Fig.~\ref{fig:jump_distr_sl} ($\tau\approx\tau_c$) multiplied by their actual system size. A fit of the distribution is shown, assuming a truncated power-law ($10^4$ bootstrap samples).}
\label{fig:jump_distr_sl_fit}
\end{figure}

The relative jump size distribution on the square lattice is illustrated in Fig.~\ref{fig:jump_distr_sl}. For $\tau\approx\tau_c$ the largest jumps at the right hand side are separated from the smaller jumps which can be seen as a kind of precursor. The separation disappears for larger values of $\tau$. This is in line with the observation made in the main texts that abrupt jumps are more likely when the transmissibility is lower. For $N$ nodes all curves begin at the minimum jump size $2 N^{-1}$, not counting the standard infection and healing processes of size $N^{-1}$. In order to analyze the pseudo-linear part for $\tau\approx\tau_c$ in Fig.~\ref{fig:jump_distr_sl} the distribution of the absolute jumps in this particular region is shown in Fig.~\ref{fig:jump_distr_sl_fit}. The data collapse to a single curve for all considered system sizes. This probability density function $f(x)$ is broad and is well fitted by a truncated power law ansatz:
\begin{equation}
f(x)\propto (x+x_0)^{-\alpha} \exp(-x/\kappa),
\end{equation}
where $x_0=-0.16\pm 0.25$, $\alpha=3.16\pm 0.29$ and $\kappa=25.67\pm 11.17$. The errors are estimated with the bootstrapping technique ($10^4$ samples).
\clearpage
\bibliography{refs}
\bibliographystyle{apsrev4-1}
\end{document}